\documentstyle[hip-artc]{article}  
\volnumber{9}  \edyear{1999}  \frompage{?} \topage{?? (submitted)}       
\recrevdate{1 March   1999}                                              
\def\rightmark{Coherent States of Creation} 
\def\leftmark{T. Cs\"org\H o}
\def\ket#1{|#1\rangle}
\def\bra#1{\langle#1|}
\def\brak#1#2{\langle#1|#2\rangle}
\def\be{\begin{equation}}
\def\ee{\end{equation}}
\def\bea{\begin{eqnarray}}
\def\eea{\end{eqnarray}}
\def\dst{\displaystyle\phantom{|}}
\def\ov{\over\displaystyle\phantom{|}}

\title{Coherent States of the Creation Operator\\
	from Fully Developed Bose-Einstein Condensates} 
\authors{ 
{\twerm T. Cs\"org\H o$^{1,2}$
}\\[2.812mm]
{\normalsize
\hspace*{-8pt}$^1$ Department of Physics, Columbia University \\ 
538 W 120-th Street, New York, NY 10028\\[0.2ex] 
\hspace*{-8pt}$^2$ MTA KFKI RMKI\\ 
H - 1525 Budapest 114, POB 49, Hungary
}}
 
\abstract{
A fully developed Bose-Einstein condensate, 
containing macroscopically large number of bosons 
can under certain conditions be considered as a {\it generalized 
vacuum state}.
Applying the annihilation operator to the condensate 
hole states can be defined. Infinite ladders of such hole states
can be considered as  {\it generalized coherent states
of the  creation operator}.
Dedicated to the memory of Professor V. N. Gribov.
}
 
\begin{document}
 
\maketitle

\section{Introduction}
Coherent states of the annihilation operator are well known in
quantum optics. They were first described by E. Schr\"odinger
as classically behaving solutions of 
the Schr\"odinger equation with a harmonic potential~\cite{schrod,nieto}.
The importance of coherent states became wildely recognized in
many branches of physics due to the works of 
Glauber~\cite{glauber}, Klauder~\cite{klauder} and Sudarshan~\cite{sudar}. 

In section~\ref{s:coh} $\,\,\,$ the definition of the 
coherent states of the annihilation operator is given and the properties
of these states are briefly summarized.
Section~\ref{s:plas} $\,\,\,\,\,$ summarizes the pion-laser model,
an exactly solvable multiboson wave-packet model that can
be solved both in the very rare gas and in the fully
condensed limiting case.
If the total available energy and thus the number of
pions in the condensate is large enough, applying the annihilation
operator to the condensate state repeatedly a ladder of holes
can be created  and by a suitable superposition of these states
a new kind of state can be defined, 
as described in Section~\ref{s:csc}$\,\,\,$.
 
\section{Coherent States\label{s:coh}}  

For the harmonic oscillator, the coherent states $\ket{\alpha}$ 
can be equivalently defined with the help of the displacement operator method,
the ladder (annihilation operator) method and the minimum uncertainty method,
see ref.~\cite{nieto} for an elegant summary.

The coherent states of the annihilation operator are solutions of the
equation
\be
	a \ket{\alpha} = \alpha \ket{\alpha},
\ee
where the annihilation and creation operators $a$ and $a^{\dagger}$
satisfy the canonical commutation relations
\be
	[a, a^{\dagger}] = 1.
\ee
For the harmonic oscillator, the above coherent states are given by the
unitary displacement operator $D(\alpha)$ acting on the ground (or vacuum)
state $\ket{0}$ as
\bea
	D(\alpha) \ket{0} & = & \mbox{\rm e}^{-|\alpha|^2/2}
		\sum_{n=0}^{\infty} {\dst \alpha^n\ov \sqrt{n!}} 
			 \ket{n} \, = \, \ket{\alpha}, 
				\label{e:ladder-a}\\
	D(\alpha) & = & \exp[\alpha a^{\dagger} - \alpha^* a] \, = \,
		\exp[-|\alpha|^2/2 ] 
		\exp[\alpha a^{\dagger}]
		\exp[\alpha^* a]
\eea
It is straightforward to show that the coherent states of the harmonic
oscillator correspond to minimum uncertainty wave-packets with
$(\Delta x)^2(\Delta p)^2 = 1/4$ that retain their Gaussian shape
during their time evolution and whose mean position and coordinate
values follow the oscillations of  classical  motion of the harmonic
oscillator.
 The coordinate space representation of these coherent states is
\be
	\brak{x}{\alpha} = \left[{\dst m \omega \ov \pi}\right]^{1/4}
		\exp\left[
		- {\dst m \omega (x - x_0)^2 \ov 2} + i p_0 x
		\right]
\ee
where $m$ is the mass of the classical particle in the harmonic
oscillator potential, $\omega$ is the frequency of the
 oscillator and $x_0$, $p_0$ correspond to the coordinate
 and momentum expectation values at the initial time $t_0$ .
 The complex eigenvalue $\alpha$ of these coherent states is
 given by
\be
	\alpha = \sqrt{\dst m \omega\ov 2} x_0 
		+ i {\dst 1 \ov \sqrt{2 m \omega}} p_0.
\ee

\section{Fully condensed limit of the pion laser model \label{s:plas}}
\def\bx{{\bf{x}}}
\def\bp{{\bf{p}}}
\def\bk{{\bf{k}}}
\def\bpi{{\bf{\pi}}}
\def\bxi{{\bf{\xi}}}
\def\bq{{\bf{q}}}
\def\br{{\bf{r}}}
\def\axd{\hat{ a}^{\dag} (\bx)}
\def\apd{\hat a^{\dag} (\bp)}
\def\ax{\hat{ a}^{} (\bx)}
\def\ap{\hat  a^{} (\bp)}
\def\psxd{\hat \Psi^{\dag} (\bx)}
\def\psx{\hat \Psi^ (\bx)}
\def\psxde{\hat \Psi^{\dag} (\bx)}
\def\psxe{\hat \Psi^ (\bx)}
\def\pdpx#1{\hat \Psi^{\dag}(\mathbf{x}_{#1},\mathbf{\pi}_{#1})\ket{0}}
\def\ri{\right)}
\def\lef{\left(}
\def\dst{\displaystyle\phantom{|}}
\def\ov{\over\dst}
\def\om{\omega}
\def\Li{\mbox{\rm Li}}
\def\gLi{\mbox{\rm gLi}}
\def\eps{\epsilon}
\def\bak{{\bf K}}
\def\dek{{\bf \Delta k}}
\newcommand{\beq}{$$}
\newcommand{\eeq}{$$}
\renewcommand{\vec}[1]{{\bf #1}}

In high energy heavy ion reactions, hundreds of identical
bosons (mostly $\pi$ mesons) can be created.
These bosons are described by rather complicated 
fields and as the density of these bosons is increased
multi-particle symmetrization effects are becoming
increasingly important. The related
possibility of Bose-Einstein condensation and the development
of partial coherence was studied recently in a large number 
of papers. Let us follow refs.~\cite{cstjz} in describing
an analytically solved multiparticle wave-packet system,
with full symmetrization and the possibility of condensation
of wave-packets to the least energetic wave-packet mode.

This solvable  model is described  by a multiparticle
density matrix
\be
\hat \rho \, = \, \sum_{n=0}^{\infty} \, {p}_n \, \hat \rho_n,
\ee
normalized to one. 
Here $ \hat \rho_n $ is the density matrix for 
events with fixed particle number $n$, which is normalized 
also to one. The probability for such an event is $ p_n $.
The multiplicity distribution is described by the set
$\left\{ {p}_n\right\}_{n=0}^{\infty}$, also normalized to 1.

The density matrix of a system with a fixed number 
of boson wave packets has the form 
\be
\hat \rho_n \, =\,  \int d\alpha_1 ... d\alpha_n
\,\,\rho_n(\alpha_1,...,\alpha_n)
\,\ket{\alpha_1,...,\alpha_n} \bra{\alpha_1,...,\alpha_n},
\ee
where $\ket{\alpha_1,...,\alpha_n}$ denote properly normalized
$n$-particle wave-packet boson states.

In Heisenberg picture, the wave packet creation operator is given as
\be
\alpha_i^{\dag}  \, = \, \int {d^3\bp \over
(\pi \sigma^2 )^{3\over 4} } \ 
	\mbox{\rm e}^{
		-{( \bp- \bpi_i)^{2}\over 2 \sigma_i^{2}}
	-  i \bxi_i (\bp - \bpi_i) } \ \apd .
\label{e:4}
\ee
The  commutator  
\be
\left[ \alpha_i, \alpha_j^{\dag} \right] \, = \,
	\brak{\alpha_i}{\alpha_j}
\ee
vanishes only in the case, when the wave packets do not overlap.
 
Here 
{$\alpha_i \, = \, (\bxi_i, \bpi_i)$} refers 
to the center of the wave-packets  
{in coordinate space and in momentum space}.
It is assumed that the widths $\sigma_i$ of the wave-packets
in momentum space and 
 the production time for each of the  wave-packets coincide.

We call the attention to the fact that although one cannot 
attribute exactly defined values for space and momentum at the same
time, one can define precisely the $ \bpi_i, \bxi_i $ parameters.

The $n$ boson states, normalized to unity, are given as
\be
\ket{\ \alpha_1, \ ...\ , \ \alpha_n} \, = \,
 {1\over \sqrt{ \displaystyle{\strut
 \sum_{\sigma^{(n)} }
\prod_{i=1}^n
\brak{\alpha_i}{\alpha_{\sigma_i}}
 } } }
\  \alpha^{\dag}_n  \ ... \
\alpha_1^{\dag} \ket{0}.
\label{e:expec2}
\ee
Here $\sigma^{(n)}$ denotes the set of all the permutations of 
the indexes $\left\{1, 2, ..., n\right\}$ 
and the subscript  $_{\sigma_i}$ denotes the index that
replaces the index $_i$ in a given permutation from $\sigma^{(n)}$. 
The normalization factor contains a sum of $ n! $ term. These terms
contain $ n$ different $\alpha_i $ parameters.

There is one special density matrix, for which one can
overcome the difficulty, related to the
non-vanishing overlap of many hundreds of wave-packets,
even in an explicit analytical manner. This density matrix is the 
product uncorrelated single particle matrices multiplied with 
a correlation factor, related to stimulated emission of wave-packets
\be
\rho_n(\alpha_1,...,\alpha_n) \, = \, {\dst 1 \ov {\cal N}{(n)}}
\lef \prod_{i=1}^n \rho_1(\alpha_i) \ri \,
\lef\sum_{\sigma^{(n)}} \prod_{k=1}^n \, 
\brak{\alpha_k}{\alpha_{\sigma_k}}
\ri .
\label{e:dtrick}
\ee
Normalization  to  one  yields ${\cal N}(n)$.

For the sake of simplicity we assume a factorizable Gaussian form
for the distribution of the parameters of the single-particle 
states:
\bea
\rho_1(\alpha)& = &\rho_x(\bxi)\, \rho_p (\bpi)\, \delta(t-t_0), \\
\rho_x(\bxi) & = &{1 \over (2 \pi R^2)^{3\over 2} }\, \exp(-\bxi^2/(2
R^2) ), \\
\rho_p(\bpi) & = &{1 \over (2 \pi m T)^{3\over 2} }\, \exp(-\bpi^2/(2 m
T) ).
\eea
These expressions are given in the frame where 
we have a  non-expanding
static source  at rest.  

A multiplicity distribution when
Bose-Einstein effects are switched {off} (denoted by $p_n^{(0)}$), 
is a {free choice} in the model. We assume independent emission,
\be
        {p}^{(0)}_n \, = \,{n_0 ^n \over n!} \exp(-n_0),
\ee
so that correlations arise only due to multiparticle Bose-Einstein
symmetrization.  This completes the specification of the model.

It has been shown in refs. ~\cite{cstjz,pratt} that the above model
features a critical density. If the boson source is sufficiently
small and cold, the overlap between the various wave-packets 
can become sufficiently large so that Bose-Einstein condensation
starts to develop as soon as $n_0$, the mean multiplicity without
symmetrization  reaches a critical value $n_c$.

In the highly condensed  $ R^2 T << 1 $  and $n_0 >> n_c$ 
Bose gas limit a kind of lasing behaviour  and an optically coherent behaviour
is obtained, which is characterized by \cite{cstjz}
the disappearance of the bump in the two-particle intensity
correlation function:
\be
	C_2(\bk_1,\bk_2) = 1
\ee

Suppose that $n_{f} = E_{tot}/m_{\pi} $ quanta are in the condensed state.
It was shown in refs.~\cite{cstjz} that the condensation
happens to the wave-packet mode with the minimal energy,
i.e. $\alpha = \alpha_0 = (0,0)$.
The density matrix of the condensate can be easily given
as the fully developed Bose-Einstein condensate corresponds
to the $T \rightarrow 0$ and the $R \rightarrow 0$ limiting case,
 when the Gaussian factors in $\rho(\alpha)$ tend to Dirac delta 
 functions.
In this particular limiting case, the density matrix of 
eq.~\ref{e:dtrick} simplifies as:
\be
	\rho_c = {\dst 1 \ov n_f!} 
		(\alpha_0^{\dagger})^{n_f}
			\ket{0}\bra{0}
		(\alpha_0)^{n_f}
\ee
which means that the fully developed Bose-Einstein condensate
corresponds to $n_f$ bosons in the same minimal wave-packet state
that is centered at the origin with zero mean momentum,
$\alpha_0 = (0,0)$.
Note also that 
\be
	\rho_c^2 = \rho_c
\ee
which implies that the fully developed Bose-Einstein condensate
is in a pure state.

An important feature of such a Bose-Einstein condensate
of massive quanta is  that it becomes impossible to add more
than $n_{f}$ pions to 
the condensate as all the available energy can be
used up by the rest mass of these bosons. 

\section{Coherent states of creation operators\label{s:csc}}
Observe, that the fully developed Bose-Einstein condensate (BEC) of the
previous section corresponds to filling a single (wave-packet) 
quantum state with macroscopic amount of quanta.
As the number of quanta in the BEC is macroscopically large,
we can treate this number first to be the infinitly large
limit of the quanta in the BEC. 
Formally, such a quantum state of the condensate can be
defined as
\be
	\ket{BEC} = 
		{\dst 1\ov \sqrt{n_f!}} (\alpha_0^{\dagger})^{n_f} \ket{0},
		\qquad (n_f >>> 1).	
\ee
In what follows, it will be irrelevant that the Bose-Einstein
condensation happened to a wave-packet mode in the pion-laser
model. The relevant essential feature of the
fully developed
Bose-Einstein condensate will be that in contains
a macroscopically large number of bosons in the same
quantum state created by certain creation operator $a^{\dagger}$
\be
	\ket{BEC}  = 
		{\dst 1\ov \sqrt{n_f!}} (a^{\dagger})^{n_f} \ket{0},
		\qquad (n_f >>> 1).	
\ee
Due to the macroscopically large number of quanta in the same state,
a large number of wave-packets can be taken away from this state. 
Due to the finite energy constraint, $n_f m = E_{tot}$,
it is impossible to add one more particle to the condensate at the
prescribed $E_{tot} $ available energy. We thus have
\be
	a^{\dagger} \ket{BEC} = 0
\ee
On the other hand, we have
\be
	a^m \ket{BEC} \ne 0 \qquad \mbox{\rm for all}\quad 0 \le m \le n_f.
\ee
Hence a Bose-Einstein condensate with macroscopically large amount of 
quanta and with a constraint that the condensate is fully developed,
can be considered as a generalized vacuum state of the creation operator,
\be
	\ket{BEC} = \ket{0}_{\dagger} ,
\ee
and generalized hole-states can be defined as removing particles from
the condensate:
\be
	\ket{n}_{\dagger} = {\dst 1\ov \sqrt{n!} } a^n \ket{0}_{\dagger}
\ee
The following calculations and equations are to be done first at
$n_f$ kept finite, and then performing the $n_f\rightarrow \infty$ 
limiting case. This corresponds to the limit of a macroscopically
large Bose-Einstein condensate.
One obtains:
\bea
	\ket{0}_{\dagger} & = & \ket{ n_f} = \ket{n_f \rightarrow \infty},\\
	\ket{1}_{\dagger} & = & \ket{n_f -1} \, = \, a \ket{0}_{\dagger},\\
	... && \nonumber \\
	\ket{j}_{\dagger} & = & \ket{n_f - j} \, = \, 
			{\dst a^j \ov \sqrt{j!}}\ket{0}_{\dagger} ,\\
	... && 		\nonumber		
\eea
	The above states can be considered as the number states
	related to the creation of $n$ holes 
	in the fully developed Bose-Einstein condensate.
	These form a ladder that is built up with the help
	of the annihilation operator. 
	The creation and annihilation operators act on these
	states as
\bea
	a \ket{n}_{\dagger}  & = & \sqrt{n + 1} \ket{n+1}_{\dagger} , \\
	a^{\dagger} \ket{n}_{\dagger} & = & \sqrt{n} \ket{n-1}_{\dagger}.
\eea
	The number operator $N_{\dagger}$ that counts the number of 
	holes is given as 
\be
	N_{\dagger}  =  a a^{\dagger}.
\ee
	With other words, the creation and annihilation operators
	change role if the ground state for our considerations
	is chosen to be the quantum state of 
	a fully developed Bose-Einstein condensate.

	If the number of quanta
	in the Bose-Einstein condensate is macroscopically large,
	($\lim n_f \rightarrow \infty$), 
	an infinite ladder can be formed from these states that is not
	bounded from below. Hence, the coherent states of the 
	creation operator can be defined as follows:
\be
	\ket{\alpha}_{\dagger}  = \exp(-|\alpha|^2/2) 
			\sum_{n = 0}^{\infty} {(\alpha^*)^n  \ov \sqrt{n!}}
				\ket{n}_{\dagger} 
\ee
	Note that the above defined coherent state is an eigenstate
	of the creation operator,
\be
	a^{\dagger} \ket{\alpha}_{\dagger}  = \alpha^* \ket{\alpha}_{\dagger}.
\ee
	It is tempting to note that the coherent states of the
	creation operator are also expressable as the action of the 
	displacement operator $D^{\dagger}(\alpha)$ 
	on the fully developed
	Bose-Einstein condensate state $\ket{BEC} = \ket{0}_{\dagger}$ as
\be
	\ket{\alpha}_{\dagger} = D^{\dagger}(\alpha) \ket{0}_{\dagger}.
\ee

	It is straightforward to generalize the results to 
	different modes characterized with a momentum ${\bf k}$.
	In that case, the state $\ket{0}_{{\bf k}+}$ has to
	be introduced as a state of  fully developed Bose-Einstein
	condensate where each boson moves with momentum ${\bf k}$.
	Such moving Bose-Einstein condensates~\cite{atom1} are frequently 
	referred to as atom lasers~\cite{atom2} in atomic physics.

\section{Interpretation and summary}
	Coherent states of the annihilation operator correspond
	to semiclassical excitations of the vacuum, that follow a
	classical equation of motion and keep their shape
	minimizing the Heisenberg uncertainty relations all the time.
	They correspond to a displaced ground state of the harmonic
	oscillators.
	 
		What is the physical interpretation of the 
		new kind of coherent states of the creation
		operator described in the present work?
		We have shown that these new states correspond
		to coherently excited holes in a fully developed
		Bose-Einstein condensate. We have found that
	displaced Bose-Einstein condensates in a 
	harmonic oscillator potential can be considered
	as generalized coherent states of the creation operator.

\section*{Acknowledgements}
The author would like to thank Professor Gribov for inspiration,
Professor M. Gyulassy for kind hospitality at the Department of
Physics at Columbia University.
Support from the following grants is greatfully acknowledged:
US - Hungarian Joint Fund MAKA/652/1998, 
OTKA T026435, T024094 
NWO - OTKA grant N25487, Department of Energy grants 
DE  - FG02-93ER40764, DE - FG02-92-ER40699, DE-AC02-76CH00016.

\section*{} 
\bigskip
\rightline{\tt What we call the beginning is often the end}
\rightline{\tt And to make an end is to make a beginning.}
\rightline{\tt The end is where we start from.}
\bigskip
\rightline{from {\it Little Gidding} by T. S. Eliot}
\bigskip
\bigskip
\bigskip
Dedicated to the memory of Volodja Gribov.
\vfill\eject

\vfill\eject
\section*{Appendix: Constraints and the ladder 
representation }  
The introduction of the coherent states of the creation
operator essentially relied on the possibility of putting 
macroscopically large, $n_f >>> 1$ amount of neutral bosons to the
same quantum state. The equations describing these states
as given in the body of the paper
are correct only to the precision given by $1 \over n_f$.
We argue in this section, that such a limitation in fact
is not generic to the coherent states of the creation operator,
but also appears when energy constraints are taken into account
in the description of the well-known coherent states of the annihilation
operator.

Let us note, that the eigenmode $n$ of
the harmonic oscillator has an energy of $E_n - E_0 = n \omega$ (after removing
the contribution of the ground or vacuum state).
Hence in the superposition given by eq.~(\ref{e:ladder-a}) modes with
arbitrarily large energy component are mixed into (but the 
weight decreases as a Poisson-tail with increasing values of $n$.)
Although the expectation value of the total energy in the coherent
states is finite, the admixture of extremely high energy components
can never be perfectly realized in any experiment as with increasing 
the energy of the included modes new physical phenomena like particle
and antiparticle creation, deviations from the harmonic shape of
the oscillator potential or other non-ideal phenomena have 
to occur.

Suppose that $E_{max}$ is the maximal available energy and
states with energy larger than $E_{max}$ are not allowed
either due to e.g. constraints from energy conservation
or due to the break-down of the harmonic approximation to
the Hamiltonian after some energy scale.
In this case modes are limited to 
$ n \le n_{f} = E_{max}/\omega$.
In case of photons in electromagnetic fields, $\omega_{\bf k} = |{\bf k}|$,
hence for sufficiently soft modes $n_{f}$ can be always made so large
that the contribution of states with $n \ge n_{f}$ to the
coherent states $\ket{\alpha}$ can be made arbitrarily small.
However, for massive bosons like bosonic atoms or mesons created
in high energy physics, $\omega_{\bf k}= \sqrt{m^2 + |{\bf k}|^2}$,
hence $n_{f} \le E_{max}/m$ for any mode of the field
which yields
\be
	\ket{\alpha} \rightarrow \ket{\alpha}_m =
		\sum_{n = 0}^{n_{f}} 
		{\dst \alpha^n (a^{\dagger})^n \over n!} \ket{0} \equiv 
		D_{n_{f}}(\alpha) \ket{0}
\ee
As each mode of a free boson field is approximately a
harmonic oscillator mode, this suggests that 
coherent states of massive bosons with finite energy constraints
can only be appoximately realized.

\end{document}